\documentclass[11pt,a4paper]{article}
\pdfoutput=1
\usepackage[margin=2.55cm]{geometry}
\usepackage{amsmath,amssymb}
\usepackage{xcolor}
\usepackage[nosort]{cite}
\usepackage{microtype}
\usepackage{bm}
\usepackage{titlesec, lmodern}
\usepackage{graphicx,mathrsfs,booktabs}
\usepackage{hyperref}
\definecolor{midblue}{rgb}{0,0,0.5}
\hypersetup{colorlinks=true,linkcolor=midblue,citecolor=green!60!black,urlcolor=midblue}

\newcommand{\dd}{\mathrm{d}}
\newcommand{\Mpl}{M_{\rm Pl}}
\newcommand{\Q}{\mathcal Q_S}

\allowdisplaybreaks[2]
\titleformat{\section}{\large\bfseries}{\thesection}{0.75em}{}
\titleformat{\subsection}{\normalsize\bfseries}{\thesubsection}{0.75em}{}
\titlespacing*{\section}{0pt}{2.1ex plus .5ex minus .2ex}{1.0ex}
\titlespacing*{\subsection}{0pt}{1.7ex plus .4ex minus .2ex}{0.7ex}

\begin{document}
	\renewcommand{\thefootnote}{\fnsymbol{footnote}}
	\vspace*{1.5em}
	\begin{center}
		{\LARGE\bfseries A Stable Vector Bounce\par}
		\vspace{2.1em}
		{\large Nils A. Nilsson$^{a,b, }\footnote{\texttt{nilsson@ibs.re.kr}}$\par}
		\vspace{0.45em}
		{\itshape
			$^a$Cosmology, Gravity, and Astroparticle Physics Group,\newline
			Center for Theoretical Physics of the Universe, Institute for Basic Science,\newline
			Daejeon 34126, Korea\\[0.3em]
			$^b$LTE, Observatoire de Paris, Universit\'e PSL, CNRS, LNE, Sorbonne Universit\'e,\newline
			61 avenue de l'Observatoire, 75014 Paris, France\par}
	\end{center}
	
	\vspace{5.0em}
	\hrule
	\vspace{1.0em}
	\noindent\textbf{Abstract.}\quad
	We construct a geodesically complete bounce in cubic generalised Proca theory with a free canonical scalar as the matter field. The temporal vector evolves along its algebraic constraint and supports the transition from contraction to expansion while matter obeys the null energy condition. This evolution permits nonzero derivative mixing at the bounce. The kinetic coefficients are positive at every finite time for all nonzero comoving wavenumbers, and short-wavelength perturbations are stable throughout the solution. Following the linear equations to both infinite-time limits also shows that local perturbations are bounded for all nonzero comoving wavenumbers.
	\vspace{1.0em}
	\hrule
	\vspace{1.0em}
	
	\renewcommand{\thefootnote}{\arabic{footnote}}
	\setcounter{footnote}{0}
	
	\vspace{0.8em}
	\vspace{1.2em}

	\newpage
	
	\section{Introduction}\label{sec:intro}
	A nonsingular bounce replaces the initial cosmological singularity with a smooth transition from contraction to expansion. It allows the expanding Universe to emerge from an earlier phase and raises the question of which gravitational degrees of freedom can support such a transition. Several proposed mechanisms with different matter content exist, each with different gravitational interactions and evolution before the bounce \cite{Novello2008,Battefeld2015}. In spatially flat general relativity, the transition requires a source that violates the null energy condition (NEC) \cite{MolinaParis1999}. Additional gravitational fields can modify the background equations so that ordinary matter continues to obey the NEC. It is known, however, that issues may appear before the bounce is even reached; for example, anisotropies can grow relative to the matter density during contraction, and the fluctuations generated in that phase must be followed through the bounce to ensure that they are stable. It was found in the context of ekpyrotic scenarios \cite{Khoury2001} that a component with equation of state $w>1$ can suppress anisotropy during contraction \cite{Erickson2004}, but a mechanism for entering expansion is still needed. That mechanism must also avoid ghosts and gradient instabilities in every propagating sector. Its behaviour at early and late times also matters, since regular evolution over a finite interval does not necessarily extend to a regular history in both time directions.
	
	Scalar-tensor theories provide a setup in which derivative interactions can support the evolution required by a bounce: Horndeski theory has second-order field equations \cite{Horndeski1974} and includes kinetic gravity braiding, where derivatives of the scalar and metric mix \cite{Deffayet2010}. These interactions permit NEC-violating evolution that a minimally coupled canonical scalar cannot support. They have been used in Galileon bounces \cite{Qiu2011,Easson2011} and in Galilean Genesis, where expansion develops from an asymptotically Minkowski state \cite{Creminelli2010}. Such constructions show how derivative interactions can support a regular transition or the onset of expansion. To assess the full history, however, stability must hold away from the transition as well, and the effective description must remain valid at all relevant scales \cite{Koehn2016}.
	In Horndeski theory, requiring the scalar gradient coefficient to remain positive gives an inequality that can be integrated over the cosmological history. Under mild assumptions about the asymptotic behaviour of the coefficients in the quadratic action, this inequality excludes complete bouncing and Genesis solutions with stable perturbations, leading to the no-go results of Refs.~\cite{Libanov2016,Kobayashi2016,Creminelli2016}. The analysis also requires a good choice of time coordinate since a coordinate choice that ceases to be valid can produce apparently singular coefficients even when the perturbations remain finite and healthy. Studies of bouncing solutions and crossings of the scalar-metric mixing coefficient have examined this distinction in for example \cite{Ijjas2016,Ijjas2018}.
	
	Several constructions exist which avoid the integrated restriction in Horndeski by changing either the gradient condition or its asymptotic assumptions. Operators in the effective field theory of cosmological perturbations and in beyond-Horndeski theories modify the scalar gradient condition \cite{Cai2017,Kolevatov2017}, and the latter setting allows stable bounces with GR-like asymptotics in both time directions \cite{Mironov2018}. Within Horndeski theory, bouncing and Genesis solutions can also avoid this issue if the relevant perturbation coefficients, including the effective Planck mass squared, vanish sufficiently rapidly in the asymptotic past \cite{Ageeva2021,Choi2025}. This vanishing can raise a strong-coupling problem, so the interaction scale must be compared with the energy scale of the evolving background. Analyses of Horndeski Genesis and bouncing models have found parameter ranges in which the interaction scale remains above the background scale \cite{Ageeva2021,Ageeva2020}.
	
	Generalised Proca (GP) can be used to extend the study of nonsingular cosmologies to vector-tensor theory and contains derivative interactions which preserve the constraints that leave three vector polarisations \cite{Heisenberg2014,Tasinato2014,Allys2016}. Although the longitudinal vector mode shares features with scalar derivative theories, the transverse modes and their interactions introduce further contributions to the perturbation equations. Furthermore, stealth vector interactions can affect perturbations even when they make no contribution to the homogeneous background \cite{DeFelicePerturbations2016}. Their role must therefore be assessed together with the scalar and tensor sectors when testing the stability of a cosmological solution. In this context, bouncing backgrounds have been constructed in vector-tensor theories within the GP class \cite{Jimenez2016}, and a later perturbation analysis identified an intrinsic vector contribution that can relax the Horndeski gradient restriction \cite{Heisenberg2018}. This mechanism motivate the search for a complete solution whose perturbations remain regular throughout its evolution, but the gradient condition does not establish that the system is stable at the nonlinear level. Furthermore, a recent analysis derived no-go conditions for bouncing solutions in GP and beyond-GP theories from the temporal-vector constraint and perturbative stability requirements \cite{Bohnenblust2025}.

	In this paper, we construct a geodesically complete bounce in the cubic GP subclass, with a free canonical matter scalar obeying the NEC. The temporal vector evolves along its algebraic constraint, and its time dependence allows for nonzero derivative mixing at the bounce. Starting from the full GP action and background equations, we derive the local bounce conditions and follow the explicit solution to both infinite-time limits, establishing timelike and null geodesic completeness. We then derive the quadratic actions in a gauge valid at the bounce. The kinetic terms are positive at every finite time for nonzero comoving wavenumber, and short-wavelength propagation is stable throughout the history. The full linear equations further give bounded local perturbations for each fixed nonzero comoving wavenumber. Finally, we estimate the interaction scale associated with the direct longitudinal cubic term. 
	
	The paper is organised as follows: in Section~\ref{sec:theory} we introduce GP and show the action and equations of motion; in Section~\ref{sec:bounce-conditions} we show the background and perturbation conditions at the bounce; in Section~\ref{sec:model} we choose a specific GP subset and construct a healthy bouncing model; finally, we conclude in Section~\ref{sec:conclusions}. We use metric signature $(-+++)$ and units $c=\hbar=1$, and $G=1/(8\pi\Mpl^2)$, where $\Mpl$ is the reduced Planck mass.
	
	\section{Generalised Proca theory}\label{sec:theory}
	The possibility of a vector-driven bounce depends on the relation between the temporal-vector constraint and the metric equations. We therefore keep the full set of GP interactions when deriving the background equations before specialising to the cubic model. We use the standard GP action with minimally coupled matter \cite{Heisenberg2014,DeFelicePerturbations2016}
	\begin{equation}
		S=\int\dd^4x\sqrt{-g}\sum_{i=2}^{6}\mathcal L_i+S_m[g_{\mu\nu},\psi_m].
		\label{eq:gp-action}
	\end{equation}
	The Lagrangians are \cite{Heisenberg2014,DeFelicePerturbations2016}
	\begin{equation}
		\begin{aligned}
			\mathcal L_2={}&G_2(X,F,Y),\label{eq:gp-l2}\\
			\mathcal L_3={}&G_3(X)\nabla_\mu A^\mu,\\
			\mathcal L_4={}&G_4(X)R+G_{4,X}\bigl[(\nabla_\mu A^\mu)^2
			-\nabla_\mu A^\nu\nabla_\nu A^\mu\bigr],\\
			\mathcal L_5={}&G_5(X)G_{\mu\nu}\nabla^\mu A^\nu
			-\frac{G_{5,X}}{6}\bigl[(\nabla_\mu A^\mu)^3
			-3(\nabla_\mu A^\mu)\nabla_\rho A^\sigma\nabla_\sigma A^\rho\\*
			&\hspace{4.1em}+2\nabla_\mu A^\nu\nabla_\nu A^\rho\nabla_\rho A^\mu\bigr]
			-g_5(X)\widetilde F^{\alpha\mu}\widetilde F^\beta{}_{\mu}\nabla_\alpha A_\beta,\\
			\mathcal L_6={}&G_6(X)L^{\mu\nu\alpha\beta}\nabla_\mu A_\nu\nabla_\alpha A_\beta
			+\frac{G_{6,X}}{2}\widetilde F^{\alpha\beta}\widetilde F^{\mu\nu}
			\nabla_\alpha A_\mu\nabla_\beta A_\nu.
		\end{aligned}
	\end{equation}
	Here $G_{\mu\nu}$ is the Einstein tensor and we use the notation $F\equiv-F_{\mu\nu}F^{\mu\nu}/4$, $Y\equiv A^\mu A^\nu F_\mu{}^\alpha F_{\nu\alpha}$. The functions $G_{3,4,5,6}$ and $g_5$ depend only on $X\equiv-A_\mu A^\mu/2$, and $G_{i,X}\equiv\partial G_i/\partial X$. The scalar invariants are formed from the vector and its field strength, while the curvature interactions also involve the dual field strength and the double dual of the Riemann tensor which we define as
	\begin{equation} \tilde{F}^{\mu\nu}=\frac12\epsilon^{\mu\nu\rho\sigma}F_{\rho\sigma},\qquad
		L^{\mu\nu\alpha\beta}=\frac14\epsilon^{\mu\nu\rho\sigma}
		\epsilon^{\alpha\beta\gamma\delta}R_{\rho\sigma\gamma\delta},
		\label{eq:gp-duals}
	\end{equation}
	where Levi-Civita tensor has $\epsilon^{0123}=1/\sqrt{-g}$.
	
	On a spatially flat Friedmann--Lema\^itre--Robertson--Walker (FLRW) backgrounde
	\begin{equation}
		\dd s^2=-N^2(t)\dd t^2+a^2(t)\delta_{ij}\dd x^i\dd x^j,
	\end{equation}
	the only homogeneous configuration is a temporal vev which gives
	\begin{equation}
		\bar{A}_\mu=(\bar{A}(t),0,0,0),\qquad X=\frac{\bar{A}^2}{2N^2},
		\label{eq:gp-background-ansatz}
	\end{equation}
	where we denote the homogeneous background vector by $\bar{A}_\mu$ and its covariant temporal component by $\bar{A}(t)=\bar{A}_0(t)$.
	To obtain the background equations of motion, we vary the minisuperspace action with respect to the lapse $N$, the scale factor $a$, and the covariant vector component $\bar{A}^\mu$, treating them as independent variables, and then set $N=1$. All subsequent background equations are written in cosmic time $X=\bar{A}^2/2$. Also, since the background field strength vanishes, the $F$ and $Y$ dependence of $G_2$, the $g_5$ interaction and $\mathcal L_6$ make no contribution to the background equations, although they generally contribute at the level of perturbations. In the background expressions below, we evaluate $G_2$ and its $X$ derivatives at $(X,F,Y)=(\bar{A}^2/2,0,0)$, and all other functions at $X=\bar{A}^2/2$.
	
	For minimally coupled matter with density $\rho_m$ and pressure $P_m$, the lapse and scale-factor equations give
	\begin{align}
		\rho_m={}&G_2-\bar{A}^2G_{2,X}+3H\bar{A}^3G_{3,X}+6H^2G_4
		-6H^2\bar{A}^2(2G_{4,X}+\bar{A}^2G_{4,XX})\notag\\
		&-H^3\bar{A}^3(5G_{5,X}+\bar{A}^2G_{5,XX}),
		\label{eq:gp-density}\\
		-P_m={}&G_2+\bar{A}^2\dot{\bar{A}}\,G_{3,X}+2G_4(3H^2+2\dot H)
		-2\bar{A}G_{4,X}(3H^2\bar{A}+2H\dot{\bar{A}}+2\bar{A}\dot H)\notag\\
		&-4H\bar{A}^3\dot{\bar{A}}\,G_{4,XX}-H^2\bar{A}^4\dot{\bar{A}}\,G_{5,XX}
		-H\bar{A}^2G_{5,X}(2\bar{A}\dot H+2H^2\bar{A}+3H\dot{\bar{A}}),
		\label{eq:gp-pressure}
	\end{align}
	and the vector equation gives
	\begin{equation}
		\bar{A} \, \mathcal C(\bar{A},H)=0,
		\label{eq:gp-vector}
	\end{equation}
	where
	\begin{equation}
		\mathcal C(\bar{A},H)=G_{2,X}-3H\bar{A} G_{3,X}
		+6H^2(G_{4,X}+\bar{A}^2G_{4,XX})
		+\bar{A}H^3(3G_{5,X}+\bar{A}^2G_{5,XX}),
		\label{eq:general-constraint}
	\end{equation}
	and is therefore a constraint. By demanding that matter is conserved gives the continuity equation as
	\begin{equation}
		\dot\rho_m+3H(\rho_m+P_m)=0.
		\label{eq:gp-conservation}
	\end{equation}
	The diffeomorphism identity relates these equations, so they are not all independent. We restrict to solutions with $\bar{A}\ne0$, for which the vector constraint in Eq.~\eqref{eq:gp-vector} requires $\mathcal C=0$. Substituting this constraint into Eq.~\eqref{eq:gp-density} simplifies the density equation to
	\begin{equation}
		\rho_m=G_2+6H^2(G_4-\bar{A}^2G_{4,X})-2H^3\bar{A}^3G_{5,X}.
		\label{eq:gp-density-branch}
	\end{equation}
	
	\section{Conditions at the bounce}\label{sec:bounce-conditions}
	We next ask how the constraint $\mathcal C=0$ affects the transition from contraction to expansion. Eq.~\eqref{eq:general-constraint} contains no $\dot{\bar{A}}$, but it relates $\bar{A}$ itself to the Hubble parameter and therefore allows $\bar{A}$ to evolve. We use the subscript $B$ for quantities evaluated at the bounce, where $H_B=0$ and $\dot H_B>0$. Assuming from now on that $G_{2,XX}(X_B)\ne0$, then $\mathcal C_{,\bar{A}}|_B=\bar{A}_BG_{2,XX}(X_B)\ne0$, so the constraint $\mathcal{C}(\bar{A},H)=0$ determines a local curve $\bar{A}=\bar{A}(H)$, and differentiating along that curve gives
	\begin{equation}
		0=\mathcal C_{,H}+\mathcal C_{,\bar{A}}\bar{A}_{,H},\quad
		\bar{A}_{,H}\big|_B=\frac{3G_{3,X}}{G_{2,XX}}\bigg|_B.
		\label{eq:chain}
	\end{equation}
	Here $\mathcal C_{,H}$ and $\mathcal C_{,\bar{A}}$ are partial derivatives, whereas $\bar{A}_{,H}=\dot{\bar{A}}/\dot H$ follows the evolution of the vector amplitude. Along the solution, this derivative contains the change in the amplitude in Eq.~\eqref{eq:chain}.
	
	To determine whether this evolution is compatible with positive kinetic terms, we use the standard GP perturbation coefficients
	\cite{DeFelicePerturbations2016,Heisenberg2018,Bohnenblust2025}.
	Expressing $w_2$ and $w_5$ through derivatives of
	$\mathcal C(\bar{A},H)$, and using the background equations
	on the branch $\mathcal C=0$, gives
	\begin{equation}
		\begin{alignedat}{3}
			&w_2 = -\frac{\bar{A}^2}{3}\mathcal C_{,H}, &\quad
			&w_5 = \frac{\bar{A}^3}{2}\mathcal C_{,\bar{A}},&\quad
			&q_S = 3w_2^2+4q_Tw_5, \\
			&\Q = \frac{q_Tq_S}{(2Hq_T+w_2)^2},&\quad
			&\rho_m+P_m = -2q_T\dot H-w_2\frac{\dot{\bar{A}}}{\bar{A}}, &\quad
			&0=-3w_2\bar{A}\dot H+2w_5\dot{\bar{A}},
		\end{alignedat}\label{eq:general-identities}
	\end{equation}
	where $q_T=2G_4-2\bar{A}^2G_{4,X}-H\bar{A}^3G_{5,X}$ is the tensor kinetic coefficient and $\Q$ is the longitudinal kinetic coefficient. The last two relations are the metric acceleration equation and the differentiated vector constraint, $\bar{A}^3 \, \dot{\mathcal C}=0$. The first two definitions measure the response of the constraint to changes in the Hubble parameter and vector amplitude, respectively. For $q_T>0$, requiring $q_S>0$ also gives $\Q>0$ wherever $2Hq_T+w_2\ne0$. For $q_S\ne0$, solving the acceleration equation and differentiated constraint for the two background time derivatives (the last two relations in Eq.~\eqref{eq:general-identities}) gives
	\begin{equation}
		\dot H=-\frac{2w_5(\rho_m+P_m)}{q_S},\quad
		\dot{\bar{A}}=-\frac{3\bar{A}w_2(\rho_m+P_m)}{q_S}.
		\label{eq:corrected-flow}
	\end{equation}
	As long as the matter enthalpy $\rho_m+P_m$ is positive, the first relation in Eq.~\eqref{eq:corrected-flow} requires $w_{5,B}<0$ to obtain a bounce with $\dot H_B>0$ and $q_{S,B}>0$. Combining this with the demand for positive tensor and longitudinal kinetic coefficients gives the interval
	\begin{equation}
		-\frac{3w_{2,B}^2}{4q_{T,B}}<w_{5,B}<0.
		\label{eq:allowed-interval}
	\end{equation}
	The lower bound keeps $q_{S,B}>0$ by requiring that the mixing contribution $3w_{2,B}^2$ exceeds the magnitude of the negative term $4q_{T,B}w_{5,B}$. The interval also automatically requires $w_{2,B}\ne0$, so $\mathcal Q_{S,B}$ is finite and positive. For matter obeying the NEC, the enthalpy must be strictly positive at the bounce; conversely, if $\rho_{m,B}+P_{m,B}$ vanishes while $q_S>0$, Eq.~\eqref{eq:corrected-flow} gives $\dot H_B=\dot{\bar{A}}_B=0$. These are necessary and sufficient local conditions on the background and kinetic coefficients for the system to be to have healthy perturbations at the time of the bounce. Once we pick a specific model, we must also check that the perturbations are regular and healthy for all times\footnote{One may also worry that the matter sector becomes ill-defined at the bounce, since the conservation equation gives $\dot\rho_m=0$ at that point, which may indicate divergent matter sound speed. However, barotropic matter also satisfies $\dot P_m=(\dd P_m/\dd\rho_m)\dot\rho_m=0$, so the ratio $\dot P_m/\dot\rho_m$ is indeterminate at that instant and does not imply a divergence of the physical sound speed. For example, a free canonical scalar has $P_m=\rho_m$ and $c_M^2=1$ through the bounce, while radiation has $c_M^2=1/3$.}.
	
	\section{Explicit model}\label{sec:model}
	Given what we discovered in the previous section, we construct an explicit example of a stable vector bounce in the cubic GP subclass with $G_4=\Mpl^2/2$, $G_5=g_5=G_6=0$, no $Y$ dependence in $G_2$, and a free canonical matter scalar $\varphi$. The covariant action is
	\begin{equation}
		\begin{split}
			S=\int\!\dd^4x\sqrt{-g}\bigg[&\frac{\Mpl^2}{2}R+G_2(X,F)+G_3(X)\nabla_\mu A^\mu-\frac12\partial_\mu\varphi\partial^\mu\varphi\bigg].
		\end{split}
		\label{eq:action}
	\end{equation}
	The matter density and pressure satisfy $\rho_m=P_m=\dot\varphi^2/2$, while the scalar equation  of motion is $\ddot\varphi+3H\dot\varphi=0$. Equations~\eqref{eq:gp-density}--\eqref{eq:general-constraint}, restricted to this subclass and to $\bar{A}\ne0$, give the vector constraint and the two Friedmann equations as
	\begin{equation}
		G_{2,X}=3H\bar{A} G_{3,X},\quad \rho_m=G_2+3\Mpl^2H^2,\quad
		2\Mpl^2\dot H+\bar{A}^2G_{3,X}\dot{\bar{A}}=-\dot\varphi^2.
		\label{eq:background}
	\end{equation}
	To obtain an explicit solution, we use a potential quadratic in $X$ and a cubic derivative interaction of the form
	\begin{equation}
		G_2(X,F)=F+V_0-\frac{\lambda_2}{2}(X-X_0)^2,\qquad G_3(X)=g_3X,
		\label{eq:general-potential}
	\end{equation}
	with all free parameters positive. Evaluating Eq.~\eqref{eq:background} at the bounce gives $\bar{A}_B^2=2X_0$ and $\rho_{m,B}=G_2(X_B,0)=V_0$. These relations fix the vector background amplitude and matter density at the bounce, while $\lambda_2$ sets the curvature of $G_2$ and $g_3$ controls the derivative mixing. 
	
	To evaluate the longitudinal kinetic coefficient and propagation speed at the bounce, we define the quantities
	\begin{equation}
		\Theta=2\Mpl^2H+w_2,\qquad
		\Q=\frac{\Mpl^2q_S}{\Theta^2},
		\label{eq:coefficients}
	\end{equation}
	and for the cubic theory considered here, with $G_{2,F}=1$ and canonical matter, the GP scalar gradient identity reads
	\cite{Heisenberg2018}
	\begin{equation}
		\frac1a\frac{\dd}{\dd t}
		\left(\frac{2a\Mpl^4}{\Theta}\right)
		=\Q c_S^2+\Mpl^2
		+\frac{2\Mpl^4\dot\varphi^2}{\Theta^2}
		-\frac{2\Mpl^4w_2^2}{\bar{A}^2\Theta^2}.
		\label{eq:gradient}
	\end{equation}
	Here, $c_S^2$ is the squared propagation speed of the longitudinal vector mode in the high-k limit. These expressions apply where $\Theta\ne0$, which holds at the bounce because $\Theta_B=w_{2,B}=g_3\bar{A}_B^3\ne0$.
	Evaluating Eqs.~\eqref{eq:general-identities}
	and \eqref{eq:gradient} at the bounce, with the time derivatives determined by Eq.~\eqref{eq:corrected-flow}, gives
	\begin{equation}
		\begin{gathered}
			\mathcal Q_{S,B}=\Mpl^2\left(3-\frac{\lambda_2\Mpl^2}{g_3^2X_0}\right),\qquad
			\lim_{V_0\to0^+}c_{S,B}^2=\frac{\Mpl^2/X_0-1}{3-\lambda_2\Mpl^2/(g_3^2X_0)}.
		\end{gathered}
		\label{eq:general-bounce-values}
	\end{equation}
	The conditions $0<\lambda_2\Mpl^2<3g_3^2X_0$ and $0<X_0<\Mpl^2$ therefore permit a bounce with a positive kinetic coefficient and squared propagation speed for sufficiently small $V_0>0$. We set $\mathcal Q_{S,B}=\Mpl^2$ and require that $c_{S,B}^2\to1/2$ in order to keep both quantities away from zero. Since $w_{2,B}=g_3\bar{A}_B^3$, defining $g>0$ by $w_{2,B}=g\Mpl^3$ fixes $X_0=2\Mpl^2/3$, $\lambda_2=9g^2/16$, and $g_3=3\sqrt3g/8$. The remaining parameter is $V_0=\rho_{m,B}$, which we express in terms of a dimensionless density $\epsilon$ through $\rho_{m,B}=\epsilon g^2\Mpl^4$. The final model functions then take the form
	\begin{equation}
		\begin{gathered}
			G_2(X,F)=F+g^2\left[\epsilon\Mpl^4-\frac{(3X-2\Mpl^2)^2}{32}\right],\quad
			G_3(X)=\frac{3\sqrt3}{8}gX,
		\end{gathered}
		\label{eq:potential}
	\end{equation}
	where $g$ is dimensionless and sets the time scale, while $\epsilon>0$ sets the bounce density. The stability inequalities at the bounce still hold under sufficiently small changes of these coefficients.
	
	We can now express the background equations of motion in terms of the vector background $\bar{A}$: substituting Eq.~\eqref{eq:potential} into the first two relations in Eq.~\eqref{eq:background} gives $H(\bar{A})$ and $\rho_m(\bar{A})$ as
	\begin{equation}
		\begin{gathered}
			H(\bar{A})=\frac{g(4\Mpl^2-3\bar{A}^2)}{12\sqrt3 \bar{A}},\quad
			\rho_m(\bar{A})=g^2\left[\epsilon\Mpl^4
			-\frac{(3\bar{A}^2-4\Mpl^2)^2(9\bar{A}^2-8\Mpl^2)}{1152\bar{A}^2}\right],
		\end{gathered}\label{eq:branch}
	\end{equation}
	with the bounce at $\bar{A}_B=2\Mpl/\sqrt3$. Substituting the model functions and $H(\bar{A})$ into Eq.~\eqref{eq:general-identities} gives
	\begin{equation}
		\begin{gathered}
			q_T=\Mpl^2,\quad w_2=\frac{3\sqrt3}{8}g\bar{A}^3,\quad q_S=\frac{3g^2\bar{A}^2}{64}(27\bar{A}^4-12\Mpl^2\bar{A}^2-16\Mpl^4),\\
			w_5=\frac{\bar{A}^4G_{2,XX}-3Hw_2}{2}
			=-\frac{3g^2\bar{A}^2(3\bar{A}^2+4\Mpl^2)}{64}.
		\end{gathered}\label{eq:model-coefficients}
	\end{equation}
	To follow this solution in time, we insert $\rho_m+P_m=2\rho_m$ into Eq.~\eqref{eq:corrected-flow} and differentiate the density in Eq.~\eqref{eq:branch}, which gives
	\begin{equation}
		\dot{\bar{A}}=-\frac{6\bar{A}w_2\rho_m}{q_S},\quad
		\frac{\dd\rho_m(\bar{A})}{\dd \bar{A}}=\frac{Hq_S}{\bar{A}w_2},
		\label{eq:branch-flow}
	\end{equation}
	both of which hold when $H=0$. Where $q_S,\rho_m>0$, $\bar{A}$ decreases with time. Since $H(\bar{A})$ decreases with $\bar{A}$, this evolution makes $H$ increase through zero while matter obeys the null energy condition. 
	
\begin{figure*}[t]
		\centering
		\includegraphics[width=0.49\textwidth]{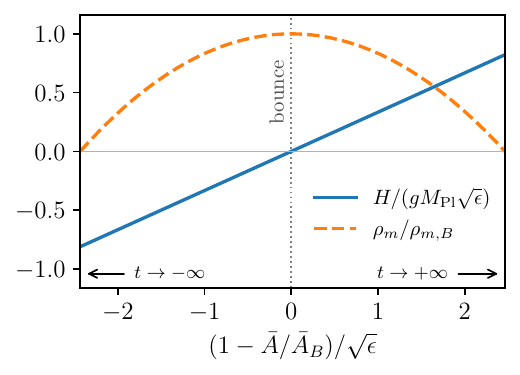}\hfill
		\includegraphics[width=0.49\textwidth]{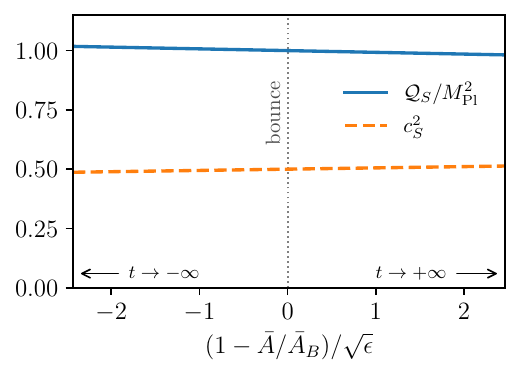}
		\caption{Bouncing solution for $\epsilon=10^{-6}$ illustrating Eq.~\eqref{eq:stability-bounds}. The coordinate $(1-\bar{A}/\bar{A}_B)/\sqrt\epsilon$ increases with cosmic time; its endpoints correspond to the infinite past and future, and the dotted line marks the bounce. \underline{Left}: Hubble parameter and matter density. \underline{Right}: Longitudinal kinetic coefficient and sound speed.}
		\label{fig:bounce}
	\end{figure*}

	To determine how far the solution extends on either side of the bounce, we first locate the zeros of the matter density. Expanding about $\bar{A}=\bar{A}_B$ gives
	\begin{equation}
		\frac{\rho_m}{g^2\Mpl^4}
		=\epsilon-\frac16\left(\frac{\bar{A}}{\bar{A}_B}-1\right)^2
		+\mathcal{O}\!\left[\left(\frac{\bar{A}}{\bar{A}_B}-1\right)^3\right].
		\label{eq:density-expansion}
	\end{equation}
	For all sufficiently small positive $\epsilon$, the density has two simple roots near $\bar{A}_B$, meaning that its first derivative is nonzero at each root. Their positions are
	\begin{equation}
		\frac{\bar{A}_\pm}{\bar{A}_B}=1\pm\sqrt{6\epsilon}+\mathcal{O}(\epsilon),
		\label{eq:root-asymptotics}
	\end{equation}
	and between these roots, $\rho_m$ is positive and reaches its maximum at the bounce as
	\begin{equation}
		0<\rho_m(\bar{A})\le\rho_{m,B}=\epsilon g^2\Mpl^4.
		\label{eq:roots}
	\end{equation}
	As $\bar{A}$ decreases through the interval $\bar{A}_-<\bar{A}<\bar{A}_+$, the solution contracts from $\bar{A}_+$, passes through $\bar{A}_B$, and expands towards $\bar{A}_-$, as shown in Fig.~\ref{fig:bounce}. 
	The endpoint values satisfy $|H(\bar{A}_\pm)|=g\Mpl\sqrt{2\epsilon/3}+\mathcal{O}(g\Mpl\epsilon)$. Decreasing $\epsilon$ therefore keeps the solution close to $\bar{A}_B$ and lowers its curvature. The restriction $\epsilon\ll1$ corresponds to a bounce density $V_0=\rho_{m,B}$ small compared with $g^2\Mpl^4$. At $\bar{A}_B$, Eqs.~\eqref{eq:corrected-flow} and \eqref{eq:model-coefficients} give $\dot H_B=2\epsilon g^2\Mpl^2>0$, $q_{S,B}=g^2\Mpl^6$, and $\mathcal Q_{S,B}=\Mpl^2$, satisfying the strict inequalities in Eq.~\eqref{eq:allowed-interval}.
	
	Finally, it remains to be checked whether the endpoints occur at finite or infinite time. Integrating the equation for $\bar{A}(t)$ in Eq.~\eqref{eq:branch-flow} expresses cosmic time as a function of $\bar{A}$, while scalar conservation gives $\rho_m a^6=\rho_{m,B}a_B^6$. The background is therefore determined by
	\begin{equation}
		t-t_B=-\int_{\bar{A}_B}^{\bar{A}}\frac{q_S(\widetilde{\bar{A}})\dd\widetilde{\bar{A}}}
		{6\widetilde{\bar{A}} w_2(\widetilde{\bar{A}})\rho_m(\widetilde{\bar{A}})},\quad
		a=a_B\left(\frac{\rho_{m,B}}{\rho_m}\right)^{1/6}.
		\label{eq:quadrature}
	\end{equation}
	At both endpoints, $\rho_m(\bar{A}_\pm)=0$ and $\left.\dd\rho_m/\dd \bar{A}\right|_{\bar{A}_\pm}\ne0$, so the density vanishes linearly. The time integral consequently diverges logarithmically as $\bar{A}\to \bar{A}_\pm$: the solution reaches these endpoints only at infinite cosmic time, with $a\to\infty$ and $H\to H_\pm\ne0$. Throughout the evolution, the scale factor remains at or above its positive minimum $a_B$.
	
	To establish geodesic completeness, we must also show that freely falling trajectories have an infinite range of proper time or affine parameter. Spatial translation symmetry gives a conserved comoving momentum of magnitude $\Pi$ along each geodesic. For a massive particle, normalising its four-velocity gives $\dd\tau/\dd t=(1+\Pi^2/a^2)^{-1/2}\ge(1+\Pi^2/a_B^2)^{-1/2}>0$, where $\tau$ is proper time and we have used $a\ge a_B$. The positive lower bound ensures that an infinite interval of cosmic time contains an infinite interval of proper time. For a light ray, the null condition gives $\dd\lambda/\dd t=a/\Pi\ge a_B/\Pi>0$, where $\lambda$ is an affine parameter and $\Pi>0$. Since cosmic time extends to both $-\infty$ and $+\infty$, integrating these bounds shows that proper time and the null affine parameter extend without bound in both directions. The spacetime is therefore timelike and null geodesically complete.
	
	\subsection{Linear cosmological perturbations}\label{sec:cosmoperts}
	We now know that the background solution is smooth through the bounce. To determine whether its perturbations are healthy, we introduce linear perturbations and expand to quadratic order. write the perturbed metric as
	\begin{equation}
		ds^2=-N^2dt^2+h_{ij}(dx^i+N^i dt)(dx^j+N^j dt),
	\end{equation}
	where $N$ is the lapse, $N^i$ is the shift, and $h_{ij}$ is the spatial metric. With $N_i=h_{ij}N^j$, we parametrise the metric and spatial vector perturbations as
	\begin{equation}
		\begin{aligned}
			N=1+\alpha, \quad
			N_i=\partial_i\beta+S_i, \quad 
			h_{ij}=a^2e^{2\zeta}(e^\gamma)_{ij}, \quad
			A_i=\partial_i\chi+V_i,
		\end{aligned}
	\end{equation}
	where $S_i$ and $V_i$ are transverse and $\gamma_{ij}$ is transverse traceless. This parametrisation has already fixed the spatial gauge freedom. We then impose the uniform vector gauge, $\chi=0$. Since a time shift acts as $\chi\to\chi-\bar{A}\delta t$, this condition fixes the remaining time-coordinate freedom provided $\bar{A}\ne0$ \cite{Heisenberg2018}. This holds throughout the solution, so the gauge remains valid at the bounce and for all values of $H$ and $\dot{\phi}$.
	
	We denote the component of the vector normal to the spatial slices by $A_\perp=(A_0-N^iA_i)/N$ and its perturbation by $\delta A_\perp$. Writing $A_0=\bar{A}+\delta A_0$ and expanding we obtain
	\begin{equation}
		\delta A_\perp=\delta A_0-\bar{A}\alpha.
		\label{eq:normal-perturbation}
	\end{equation}
	This temporal perturbation, along with the lapse $\alpha$ and the scalar shift $\beta$ are nondynamical and we integrate them out, which leaves $\zeta$ and $\delta\varphi$ as the two propagating scalars modes. The coefficients $\Theta$ and $\Q$ describing the background mixing are defined in Eq.~\eqref{eq:coefficients}. At the bounce, $q_S=g^2\Mpl^6$ and $\Theta=g\Mpl^3$. The background interval between the two density roots described above shrinks towards $\bar{A}_B$ as $\epsilon\to0^+$, so both coefficients remain positive throughout the solution for sufficiently small $\epsilon$.

	We write the quadratic actions in Fourier space using $k$ the comoving wavenumber. After integrating out the nondynamical lapse, shift, and temporal vector perturbation, the quadratic scalar action reads
	\begin{equation}
		\begin{aligned}
			S_S^{(2)}={}&
			\frac12\int\!\dd t\,\frac{\dd^3k}{(2\pi)^3}\,a^3
			\Bigg[
			6\Mpl^2\dot\zeta^2+\delta\dot\varphi^2
			+\frac{k^2}{a^2}
			\left(2\Mpl^2\zeta^2-\delta\varphi^2\right)\\&
			\hspace{7em}+\frac{1}{w_2^2}\left(2w_5+\bar{A}^2\frac{k^2}{a^2}\right)\mathcal U^2
			-\frac{\frac{1}{w_2^2}
				\left(\mathcal V_0+\frac{k^2}{a^2}\mathcal V_2\right)^2
			}{
				w_2^2\dot\varphi^2+2\Mpl^2H^2q_S
				+\bar{A}^2\Theta^2\frac{k^2}{a^2}
			}\Bigg]
		\end{aligned}
		\label{eq:scalar-action}
	\end{equation}
	where $\mathcal U = 2\Mpl^2\dot\zeta+\dot\varphi\,\delta\varphi$, $\mathcal V_0 = Hq_S\mathcal U+w_2^2\dot\varphi\,\delta\dot\varphi$, and $\mathcal V_2 = \bar{A}^2\Theta\,\mathcal U-2\Mpl^2w_2^2\zeta$.
	The denominator in the final term is positive at every finite time for $k>0$, including at $H=0$, so eliminating the constraints introduces no singularity at the bounce. Eliminating the nondynamical fields produces the factor of $k^4$ in the numerator, but the denominator also depends on $k^2$. Their combination leaves the leading high-k gradient terms proportional to $k^2$, as we show below.
	
	To check the signs of the scalar kinetic terms at each nonzero $k$, we form the kinetic matrix $K$ of the reduced quadratic Lagrangian with respect to the propagating fields $(\dot\zeta,\delta\dot\varphi)$. The matter diagonal  and determinant are
	\begin{equation}
		\begin{aligned}
			K_{\varphi\varphi}
			&=1-\frac{w_2^2\dot\varphi^2}
			{\mathcal D_0+\bar{A}^2\Theta^2\frac{k^2}{a^2}}>0,\quad 
			\det K
			=\frac{k^2}{a^2}\frac{2\Mpl^2\bar{A}^2q_S}
			{\mathcal D_0+\bar{A}^2\Theta^2\frac{k^2}{a^2}}>0,
		\end{aligned}
		\label{eq:kinetic}
	\end{equation}
	where $\mathcal D_0\equiv w_2^2\dot\varphi^2+2\Mpl^2H^2q_S$. These two positive minors establish that the kinetic matrix has two positive eigenvalues at every finite time and $k>0$.
	
	The propagation speeds follow from the leading terms at short wavelengths. We make the field redefinition $\delta\varphi_c=\delta\varphi-2\Mpl^2\dot\varphi\zeta/\Theta$ and integrate by parts to write the leading terms with two derivatives as
	\begin{equation}
		\begin{aligned}
			S_{S}^{(2)}
			\simeq\int\!\dd t\,\frac{\dd^3k}{(2\pi)^3}\,a^3
			\left[
			\Q\left(
			\dot\zeta^2-\frac{c_S^2k^2}{a^2}\zeta^2\right)
			+\frac12\left(
			\delta\dot\varphi_{c}^2
			-\frac{k^2}{a^2}\delta\varphi_{c}^2
			\right)
			\right],
			\label{eq:principal-action}
		\end{aligned}
	\end{equation}
	from which we see that the matter mode propagates with unit speed, while $c_S^2$ is the squared speed of the longitudinal vector mode. Before inserting the model functions, it is given by $c_S^2=[2\Mpl^2(H\Theta-\dot\Theta)-\Theta^2+2\Mpl^2\left(w_2^2/\bar{A}^2-\dot\varphi^2\right)]/q_S$. Using the background equations to evaluate this ratio of gradient and kinetic coefficients gives
	\begin{equation}
		c_S^2=-\frac13+
		\frac{g^2\bar{A}^2\Mpl^2(27\bar{A}^2-16\Mpl^2)}{32q_S}
		+\frac{24\Mpl^2w_2^2\rho_m}{q_S^2}.
		\label{eq:speeds}
	\end{equation}
	To establish that the kinetic and gradient coefficients stay positive throughout the evolution, we use the same expansion that describes the background near the bounce. Equation~\eqref{eq:root-asymptotics} gives $\bar{A}/\bar{A}_B=1+\mathcal{O}(\sqrt\epsilon)$ over the whole solution, while Eq.~\eqref{eq:roots} bounds the normalised matter density by $\epsilon$. The coefficients are smooth, and their denominators remain nonzero near $\bar{A}_B$. Expanding $\Q$ and $c_S^2$ in Eqs.~\eqref{eq:coefficients} and \eqref{eq:speeds} about $\bar{A}=\bar{A}_B$, with $|\bar{A}-\bar{A}_B|/\bar{A}_B=\mathcal O(\sqrt\epsilon)$ throughout the interval therefore gives the estimates
	\begin{equation}
		\frac{\Q}{\Mpl^2}=1+\mathcal{O}(\sqrt\epsilon),\quad
		c_S^2=\frac12+\mathcal{O}(\sqrt\epsilon).
		\label{eq:stability-bounds}
	\end{equation}
	These estimates establish that $\Q>0$ and $0<c_S^2<1$ for all sufficiently small positive $\epsilon$, including at both endpoints. At the bounce, the exact squared speed is $c_{S,B}^2=1/2+24\epsilon$. Since the matter density roots are simple and the stability inequalities are strict, these properties hold under sufficiently small changes of the free parameters in Eq.~\eqref{eq:general-potential}.
	
	The difference from the Horndeski gradient constraint can be read from the final term of Eq.~\eqref{eq:gradient}.
	The background field strength vanishes, but scalar perturbations of the vector carry an electric field of the form $\delta F_{0i}=\partial_i(\dot\chi-\delta A_0)$. When we integrate out the nondynamical fields from the quadratic action, this electric contribution produces the final term in Eq.~\eqref{eq:gradient}. Its negative sign removes the positive lower bound used in the Horndeski argument. This mechanism was identified in Ref.~\cite{Heisenberg2018}. The complete background constructed here also allows us to follow the full linear evolution.
	
	To complete the quadratic analysis, we write the tensor and vector actions which after integration by parts read
	\begin{align}
		S_T^{(2)}&=\frac{\Mpl^2}{8}\int\!\dd t\,\frac{\dd^3k}{(2\pi)^3}\,a^3
		\left[\dot\gamma_{ij}^2-\frac{k^2}{a^2}\gamma_{ij}^2\right],
		\label{eq:tensor-action}\\
		S_V^{(2)}&=\frac12\int\!\dd t\,\frac{\dd^3k}{(2\pi)^3}\,a
		\left[\dot V_i^2-\left(\frac{k^2}{a^2}
		+\frac{6w_2^2\rho_m}{\bar{A}^2q_S}\right)V_i^2\right],
		\label{eq:vector-action}
	\end{align}
	both of which have positive kinetic terms and unit propagation speeds in the high-k limit. The vector mass squared, $-G_{3,X}\dot{\bar{A}}=6w_2^2\rho_m/(\bar{A}^2q_S)$, is positive at finite times and tends to zero at the vacuum endpoints. Together with the scalar action, these expressions describe two tensor polarisations, three Proca polarisations, and the matter scalar, leaving $2+3+1$ propagating degrees of freedom.
	
	We can now relate these results to the perturbation coefficients introduced in Ref.~\cite{Bohnenblust2025}. The definitions in Appendix~A of that reference give for $H\ne0$ that
	\begin{equation}
		\alpha_3=\frac{w_6+w_2/\bar{A}}{4H}
		=-2\bar{A}G_{4,X}-H\bar{A}^2G_{5,X}.
		\label{eq:alpha3-regular}
	\end{equation}
	For smooth couplings in generalised Proca theory, the expression on the right-hand side extends smoothly through $H=0$. In the cubic model studied here, $\alpha_3=0$, and Eq.~(3.40) of Ref.~\cite{Bohnenblust2025} gives
	\begin{equation}
		E_3=\frac{\Mpl^2w_2}{\bar{A}\Theta},\qquad
		E_{3,B}=\frac{\Mpl^2}{\bar{A}_B}.
		\label{eq:E3-regular}
	\end{equation}
	Since $\bar{A}$ and $\Theta$ remain nonzero near the bounce, $E_3$ is regular there as well. Furthermore, in the notation of the stability conditions (4.1) of Ref.~\cite{Bohnenblust2025}, we obtain
	\begin{equation}
		q_T=F_T=\Mpl^2,\quad q_V=F_V=1, \quad
		F_S=\Q c_S^2>0,\quad c_M^2=1.
		\label{eq:stability-checklist}
	\end{equation}
	Together with $q_S>0$ and $\rho_m+P_m=2\rho_m\ge0$, these results satisfy all necessary conditions throughout the solution for sufficiently small positive $\epsilon$, by the estimates in Eq.~\eqref{eq:stability-bounds}. The matter enthalpy is also positive at every finite time and tends to zero at both asymptotic endpoints.
	
	\subsection{Complete linear evolution}\label{sec:completelinear}
	The short-wavelength conditions establish that the kinetic and gradient terms are positive and nonzero, but following a comoving mode through the complete history of its evolution also requires knowing its behaviour at long wavelengths. For our bouncing solution, the scale factor grows without bound in both time directions, so every $k>0$ has $k/a\to0$ in either infinite-time limit. We therefore study the perturbations near the two asymptotic de Sitter regions and then connect these solutions through the bounce.
	For the endpoint calculation, we use Newtonian gauge and write $\zeta_N=\psi$ and $A_{i,N}=\partial_i\chi_N$. Neither the matter scalar nor the vector sector produces scalar anisotropic stress at linear order on this background. The scalar part of the trace-free spatial metric equations derived from Eq.~\eqref{eq:action} therefore reduces, in Fourier space, to
	$
	\left(k_i k_j-\delta_{ij}k^2/3\right)
	(\alpha_N+\psi)=0.
	$
	For $k>0$, this gives $\alpha_N=-\psi$, leaving one independent scalar metric perturbation. Substituting this relation into the remaining linear constraints and using the background equations, we obtain the algebraic system
	\begin{equation}
		\begin{pmatrix}
			\displaystyle
			\frac{k^2}{a^2}+\frac{q_S}{2\Mpl^2 \bar{A}^2}
			&
			\displaystyle
			-\frac{k^2}{a^2}\bar{A}
			\\[8pt]
			\displaystyle
			-\frac{q_S}{2\Mpl^2 \bar{A}}
			&
			\displaystyle
			2\Mpl^2\frac{k^2}{a^2}-\dot\varphi^2
		\end{pmatrix}
		\begin{pmatrix}
			\delta A_{\perp,N}\\[4pt]
			\psi
		\end{pmatrix}
		=
		\begin{pmatrix}
			\displaystyle
			\frac{k^2}{a^2}
			\left(\dot\chi_N+\frac{w_2}{\bar{A}^2}\chi_N\right)
			-\frac{3w_2\dot\varphi}{2\Mpl^2 \bar{A}}\delta\varphi_N
			\\[10pt]
			\displaystyle
			\dot\varphi\,\delta\dot\varphi_N
			+\frac{3\Theta\dot\varphi}{2\Mpl^2}\delta\varphi_N
			-\frac{w_2}{\bar{A}}\frac{k^2}{a^2}\chi_N
		\end{pmatrix}.
		\label{eq:endpoint-system}
	\end{equation}
	In what follows, we normalise the scale factor to unity at the bounce, so we have $a_B=1$.
	The two constraints above allow us to express $\delta A_{\perp,N}$ and $\psi$ in terms of $\chi_N$, $\delta\varphi_N$, and their first derivatives wherever the coefficient matrix is invertible. To identify where this reduction can be used, we compute its determinant as
	\begin{equation}
		\Delta
		=2\Mpl^2\frac{k^4}{a^4}
		+\frac{k^2}{a^2}
		\left[
		\frac{q_S(2\Mpl^2-\bar{A}^2)}{2\Mpl^2 \bar{A}^2}
		-\dot\varphi^2
		\right]
		-\frac{q_S\dot\varphi^2}{2\Mpl^2 \bar{A}^2}.
	\end{equation}
	We denote the endpoint under consideration by $\bar{A}_*=\bar{A}_+$ or $\bar{A}_-$, with the corresponding Hubble parameter being $H_*$. As $a\to\infty$, the leading contribution to the determinant is $(k^2/a^2)q_S(\bar{A}_*)(2\Mpl^2-\bar{A}_*^2)/(2\Mpl^2\bar{A}_*^2)>0$, which allows us to solve the constraints sufficiently near either endpoint. At the bounce itself, $\Delta_B<0$ for $k<k_*\simeq\sqrt{3\epsilon}g\Mpl$, so the determinant must pass through zero between the bounce and the endpoints for these values of $k$. When the determinant vanishes, the two constraints no longer determine $\delta A_{\perp,N}$ and $\psi$ uniquely. We therefore solve for these variables only near the endpoints, where the determinant is nonzero. We establish regular evolution through the intervening region below.
	
	To construct the endpoint solutions beyond their leading behaviour, we need an expansion that can be followed order by order as the scale factor grows. Matter conservation gives $\rho_m=\rho_{m,B}/a^6$ and $\dot\varphi=\dot\varphi_B/a^3$. Since the density root has a nonzero derivative, $\rho_{m,\bar{A}}(\bar{A}_*)\ne0$, we can invert the density relation near the endpoint to obtain $(\bar{A},H)=(\bar{A}_*,H_*)+\mathcal O(a^{-6})$, with analytic dependence on $1/a$, so the background therefore admits an expansion in inverse powers of the scale factor. Solving the perturbation constraints requires division by their determinant, so we must check that this operation preserves the expansion. After factoring $k^2/a^2$ out of the determinant, the remaining factor tends to the positive value $q_S(\bar{A}_*)(2\Mpl^2-\bar{A}_*^2)/(2\Mpl^2\bar{A}_*^2)>0$. The matter contribution satisfies $\dot\varphi/(k^2/a^2)=\dot\varphi_B/(k^2a)$, and hence the coefficients in the solved constraints remain analytic in $1/a$.
	
	For $k>0$, the longitudinal part of the linearised spatial vector equation obtained from the action~\eqref{eq:action} reads
	\begin{equation}
		\begin{aligned}
			\ddot\chi_N+H\dot\chi_N
			-\frac{w_2\dot{\bar{A}}}{\bar{A}^3}\chi_N
			={}\delta\dot A_{\perp,N}
			+\left(H-\frac{w_2}{\bar{A}^2}\right)\delta A_{\perp,N}
			-\bar{A}\dot\psi
			-\left(\dot{\bar{A}}+H\bar{A}\right)\psi .
		\end{aligned}
		\label{eq:longitudinal-evolution}
	\end{equation}
	Here we have used $\alpha_N=-\psi$, the background vector constraint in Eq.~\eqref{eq:background}, and $w_2=\bar{A}^3G_{3,X}$. Expanding the matter-scalar equation from the same action and using $\ddot\varphi+3H\dot\varphi=0$ gives
	\begin{equation}
		\delta\ddot\varphi_N+3H\delta\dot\varphi_N
		+\frac{k^2}{a^2}\delta\varphi_N
		=-4\dot\varphi\,\dot\psi .
		\label{eq:matter-evolution}
	\end{equation}
	These equations have not yet been expanded near the endpoints. We solve Eq.~\eqref{eq:endpoint-system} for $\delta A_{\perp,N}$ and $\psi$ and substitute these solutions into Eqs.~\eqref{eq:longitudinal-evolution} and \eqref{eq:matter-evolution}. The resulting coefficient matrix for $(\ddot\chi_N,\delta\ddot\varphi_N)$ has a determinant that tends to $2\Mpl^2/(2\Mpl^2-\bar{A}_*^2)\ne0$, so both second derivatives can be isolated without losing the analytic dependence near either endpoint. The leading equations are
	\begin{equation}
		\ddot\chi_N+H_*\dot\chi_N=0,\qquad
		\delta\ddot\varphi_N+3H_*\delta\dot\varphi_N=0.
		\label{eq:endpoint-leading}
	\end{equation}
	The longitudinal equation has a constant solution and a solution proportional to $1/a$, while the matter equation has a constant solution and a solution proportional to $1/a^3$. We next include the subleading terms in the perturbation equations to determine the corrections to these four solutions. The first corrections to the longitudinal equation multiply $\chi_N,\dot\chi_N$ by coefficients of order $k^2/a^2$ and the matter perturbations by coefficients of order $a^{-3}$. In the matter equation, retaining the leading gradient and coupling terms gives
	\begin{equation}
		\delta\ddot\varphi_N+3H_*\delta\dot\varphi_N
		+\frac{k^2}{a^2}\delta\varphi_N
		=\frac{4\bar{A}_*H_*\dot\varphi_B}{2\Mpl^2-\bar{A}_*^2}
		\frac{\dot\chi_N}{a^3}+\cdots,
		\label{eq:endpoint-matter}
	\end{equation}
	where the omitted coefficient functions are analytic and begin at order $a^{-4}$.
	
	We construct the four solutions by expanding both fields in powers of $1/a$. At each order, the terms already determined at lower orders supply the source for the new coefficients. Acting on a term $a^{-j}$, the leading longitudinal and matter operators give the factors $H_*^2j(j-1)$ and $H_*^2j(j-3)$, respectively. Where one of these factors vanishes, the corresponding source must also vanish for the expansion to continue without logarithmic terms. At $j=0$, both equations leave the constant coefficients $C_\chi,C_\varphi$ independent. At $j=1$, the longitudinal equation has no source and leaves $C_1$ free, while the corresponding matter coefficient vanishes. The absence of this matter coefficient means that the gradient term supplies no source at $j=3$. The coupling to $\dot\chi_N$ also gives no source there, since differentiating the constant longitudinal term gives zero. Thus $C_3$ remains independent, and the equations require no logarithmic terms. All other coefficients at positive orders are fixed by these four constants.
	
	To establish that the series describe actual solutions near the endpoints, we must also show that they converge. Using $\dd/\dd t=-(H/a)\dd/\dd(1/a)$, we write the equations for the fields and their time derivatives as a first-order system. In this form, $1/a$ multiplies the derivative and the coefficient matrix is analytic at $1/a=0$. The Frobenius theorem then guarantees convergence of all four power series sufficiently near either endpoint. Reconstructing $\psi$ and $\delta A_{\perp,N}$ from Eq.~\eqref{eq:endpoint-system} gives
	\begin{equation}
		\begin{aligned}
			\chi_N&=C_\chi+\frac{C_1}{a}+\mathcal{O}(a^{-2}), \quad \psi=\frac{H_*}{2\Mpl^2-\bar{A}_*^2}
			\left(-\bar{A}_*C_1+\frac{3\dot\varphi_B C_\varphi}{k^2}\right)
			\frac1a+\mathcal{O}(a^{-2}),\\
			\delta\varphi_N&=C_\varphi\left(1+\frac{1}{2H_*^2}
			\frac{k^2}{a^2}\right)
			+\frac{C_3}{a^3}+\mathcal{O}(a^{-4}),\quad
			\delta A_{\perp,N}=\mathcal{O}(a^{-2}).
		\end{aligned}
		\label{eq:endpoint-modes}
	\end{equation}
	The four constants are independent for each $k>0$. To relate these solutions to local measurements, we evaluate the electric field and the electric part of the Weyl tensor which describes tidal curvature. Using Eq.~\eqref{eq:normal-perturbation} and $\alpha_N=-\psi$, their scalar perturbations in the orthonormal frame of observers normal to the Newtonian-gauge constant-time hypersurfaces are
	\begin{equation}
		\begin{aligned}
			\delta F_{\hat0\hat i}
			&=\frac{ik_i}{a}
			\left(\dot\chi_N-\delta A_{\perp,N}+\bar{A}\psi\right)
			=\mathcal O(a^{-2}),\\
			\delta C_{\hat0\hat i\hat0\hat j}
			&=\frac{1}{a^2}
			\left(k_i k_j-\frac{\delta_{ij}k^2}{3}\right)\psi
			=\mathcal O(a^{-3}),
		\end{aligned}
		\label{eq:endpoint-observables}
	\end{equation}
	where $C_{\mu\nu\rho\sigma}$ is the Weyl tensor and hats denote orthonormal-frame components. Both perturbations are gauge invariant because the background field strength and Weyl tensor vanish. The decay rates follow directly from Eq.~\eqref{eq:endpoint-modes}, and although $\chi_N$ and $\delta\varphi_N$ can approach constants, their physical spatial gradients $(k/a)\chi_N$ and $(k/a)\delta\varphi_N$ decay. The constant matter amplitude produces no nondecaying stress-energy perturbation because the matter scalar has no potential. In Newtonian gauge, its density and momentum perturbations are
	\begin{equation}
		\delta\rho_{m,N}
		=\dot\varphi\left(\delta\dot\varphi_N+\dot\varphi\psi\right)
		=\mathcal O(a^{-5}),\qquad
		\delta T^{(m)}_{\hat0\hat i}
		=\frac{ik_i}{a}\dot\varphi\,\delta\varphi_N
		=\mathcal O(a^{-4}),
		\label{eq:endpoint-matter-stress}
	\end{equation}
	where we have used $\dot\varphi=\dot\varphi_B/a^3$ and the matter solution in Eq.~\eqref{eq:endpoint-modes}.
	The tensor and transverse-vector sectors have the same qualitative behaviour at the endpoints. Equations~\eqref{eq:tensor-action} and \eqref{eq:vector-action} give tensor solutions that are constant or proportional to $a^{-3}$, and transverse-vector solutions that are constant or proportional to $a^{-1}$. Their shear and field strengths decay.
	
	To connect the two endpoint regions through the bounce, we use the uniform vector gauge, $\chi=0$, and the reduced action derived above. For $k>0$, the denominator in Eq.~\eqref{eq:scalar-action} is positive and the kinetic matrix in Eq.~\eqref{eq:kinetic} is positive definite at every finite time, so the evolution equations for $(\zeta,\delta\varphi)$ are regular on every finite time interval. We must also check that the eliminated fields can be reconstructed without introducing a singularity. Solving the linear lapse, shift and temporal-vector constraints from Eq.~\eqref{eq:action} and using the background equations gives
	\begin{equation}
		\begin{aligned}
			\alpha&=
			\frac{\mathcal V_0+\dfrac{k^2}{a^2}\mathcal V_2}
			{w_2^2\dot\varphi^2+2\Mpl^2H^2q_S
				+\bar{A}^2\Theta^2\dfrac{k^2}{a^2}},\quad
			\delta A_\perp=\frac{\bar{A}}{w_2}
			\left(2\Mpl^2H\alpha-\mathcal U\right),\\
			\beta&=\frac{a^2}{k^2\Theta}
			\left[
			\dot\varphi\,\delta\dot\varphi
			+\frac{3\Theta\dot\varphi}{2\Mpl^2}\delta\varphi
			-2\Mpl^2\frac{k^2}{a^2}\zeta
			+\frac{q_S}{2\Mpl^2\bar{A}}\delta A_\perp
			-\dot\varphi^2\alpha
			\right],
		\end{aligned}
		\label{eq:auxiliary-reconstruction}
	\end{equation}
	where $\mathcal U$, $\mathcal V_0$ and $\mathcal V_2$ are defined below Eq.~\eqref{eq:scalar-action}. The denominator of $\alpha$ is the same positive expression that appears in that action. Since $\bar{A}$, $w_2$ and $\Theta$ remain nonzero, all three constraint solutions are regular at every finite time for $k>0$, including at $H=0$. The time shift $\delta t = -\beta$ then relates these solutions to the Newtonian-gauge variables through
	\begin{equation}
		\chi_N=\bar{A}\beta, \quad \psi=\zeta+H\beta, \quad \delta\varphi_N=\delta\varphi+\dot{\varphi}\beta, \quad \delta A_{\perp,N}=\delta A_\perp+\dot{\bar{A}}\beta,
	\end{equation}
	which are regular on the same intervals. Near either endpoint, these solutions can be expanded in the basis of Eq.~\eqref{eq:endpoint-modes}, so we can join the regular evolution over finite time intervals to the convergent endpoint expansions, which establishes that local perturbations remain bounded.
	
	As an aside, we mention that the kinetic matrix also explains why the high-k analysis alone cannot determine this behaviour. On the endpoint backgrounds, it becomes
	\begin{equation}
		K_{\varphi\varphi}=1,\quad K_{\zeta\varphi}=0,\quad
		K_{\zeta\zeta}=
		\dfrac{k^2}{a^2}\frac{2\Q \bar{A}_*^2}
		{2H_*^2\Q+\bar{A}_*^2\dfrac{k^2}{a^2}},
		\label{eq:vacuum-kinetic}
	\end{equation}
	so $K_{\zeta\zeta}$ approaches $2\Q$ at high physical momentum, but is proportional to $k^2/a^2$ as $a\to\infty$ at fixed $k$. The two limits are sensitive to different regimes of the same solution. Also, a fractional quantity such as $\delta\rho_m/\rho_m=\mathcal{O}(a)$ can grow because the background density vanishes faster than its perturbation.
	
	Exactly homogeneous perturbations ($k=0$) need to be treated separately, since constraint reduction above involves dividing by powers of $k$. To that end, we impose cosmic time by setting $\alpha=0$. Linearising the two background relations in Eq.~\eqref{eq:branch} determines $\delta H$ and $\delta\rho_m$ in terms of $\delta A_\perp$. Linearising the autonomous evolution equation in Eq.~\eqref{eq:branch-flow} then gives
	\begin{equation}
		\delta H=\frac{\dd H}{\dd\bar{A}}\delta A_\perp,\quad
		\delta\rho_m=\frac{\dd\rho_m}{\dd\bar{A}}\delta A_\perp,\quad
		\frac{\dd}{\dd t}\left(\frac{\delta A_\perp}{\dot{\bar{A}}}\right)=0.
		\label{eq:homogeneous-completeness}
	\end{equation}
	The last relation is valid at every finite time because $\dot{\bar{A}}<0$ there, and $\delta A_\perp=-\dot{\bar{A}}\Delta t$ for a constant $\Delta t$. Integrating $\dot\zeta=\delta H$ and the matter relation $\dot\varphi\,\delta\dot\varphi=\delta\rho_m$ gives the complete homogeneous scalar solution as
	\begin{equation}
		\delta A_\perp=-\dot{\bar{A}}\,\Delta t,\quad
		\delta\rho_m=-\dot\rho_m\,\Delta t,\quad
		\zeta=\lambda-H\Delta t,\quad
		\delta\varphi=c-\dot\varphi\Delta t.
		\label{eq:homogeneous}
	\end{equation}
	Here $\Delta t$, $\lambda$, and $c$ shift the time origin, spatial normalisation, and scalar zero point, respectively. For the homogeneous tensors, Eq.~\eqref{eq:tensor-action} gives $a^3\dot\gamma_{ij}=\mathrm{constant}$, so each solution is a constant plus a multiple of $\int_{t_B}^{t}\dd t'/a^3(t')$. This integral stays finite because $a$ has a positive minimum and grows exponentially at both ends. For the homogeneous spatial Proca modes, the mass term in Eq.~\eqref{eq:vector-action} decays as $a^{-6}$, giving the same constant and $a^{-1}$ endpoint behaviours as above. Their equations are regular at finite times, so these modes are also bounded. This treatment follows directly from the homogeneous equations and therefore requires no division by $k^2$.
	
	\subsection{Perturbative interactions}\label{sec:interactions}
	We next estimate the scales associated interactions, and we provide a first assessment of nonlinear perturbative validity; we while leave a complete analysis of the coupled system for future work. 
	
	We begin with the longitudinal cubic interaction and express its coefficient in terms of the canonically normalised field. At frequencies $E\sim k/a\gg g\Mpl$, we expand the linear constraints for the lapse, scalar shift and temporal vector, obtained from Eq.~\eqref{eq:action}, in $g\Mpl/E$. We then combine their leading solution with the transformation from Newtonian gauge to the uniform vector gauge to obtain $\zeta\simeq-\Theta\chi_N/(2\Mpl^2\bar{A})$. Inserting this relation into the longitudinal kinetic term in Eq.~\eqref{eq:principal-action} lets us identify the canonically normalised field as
	\begin{equation}
		\pi_c=\sqrt{\frac{q_S}{2\bar{A}^2\Mpl^2}}\,\chi_N.
		\label{eq:canonical-longitudinal}
	\end{equation}
	To extract the direct longitudinal contribution from $G_3$, we hold the background and metric fixed locally and use local inertial derivatives. To leading order in $g\Mpl/E$, the temporal-vector constraint gives $\delta A_\mu\simeq\partial_\mu\chi_N$, so expanding the $G_3$ interaction in the canonically normalised field gives
	\begin{equation}
		\mathcal L^{(3)}_{G_3,\mathrm{long}}
		=-\frac{(\partial\pi_c)^2\Box\pi_c}{\Lambda_3^3},
		\quad
		\Lambda_3^3=
		\frac{2}{G_{3,X}}
		\left(\frac{q_S}{2\bar{A}^2\Mpl^2}\right)^{3/2}
		\sim g^2\Mpl^3.
		\label{eq:cubic-scale}
	\end{equation}
	The background evolves throughout the solution, so we must also follow the normalisation and cubic coefficient away from the bounce. Since $\bar{A}/\bar{A}_B=1+\mathcal{O}(\sqrt\epsilon)$ over the whole solution, we expand in $\epsilon$ and obtain the estimates
	\begin{equation}
		\frac{q_S}{2\bar{A}^2\Mpl^2}
		=\frac38g^2\Mpl^2[1+\mathcal{O}(\sqrt\epsilon)],
		\quad
		\Lambda_3^3
		=\frac{g^2\Mpl^3}{\sqrt2}
		[1+\mathcal{O}(\sqrt\epsilon)].
		\label{eq:interaction-bounds}
	\end{equation}
	The first relation shows that the longitudinal normalisation remains finite and positive, including in both infinite-time limits. The logarithmic time derivatives of $\bar{A}$ and of the longitudinal normalisation are $\mathcal{O}(\epsilon g\Mpl)$. Both quantities therefore vary slowly over the local time intervals used in the interaction estimate. At the bounce, the scale reduces to $\Lambda_{3,B}^3=g^2\Mpl^3/\sqrt2$, while the second relation in Eq.~\eqref{eq:interaction-bounds} gives $\Lambda_3\sim g^{2/3}\Mpl$ throughout the solution. For sufficiently small $g$, the scale associated with this interaction lies above the background curvature scales $|H|,\sqrt{|\dot H|}=\mathcal O(g\sqrt\epsilon\,\Mpl)$.
	
	The coupling of the longitudinal mode to the tensor, vector and matter perturbations must also be taken into account when determining the range of perturbative validity. At cubic order, we can eliminate the nondynamical fields by substituting the linear constraint solutions into the third-order action. Terms containing second-order corrections to those fields are proportional to the linear constraints and therefore vanish after the linear solutions have been imposed.
	We examine a subset of covariant operators by expanding them in local inertial coordinates and inserting the leading linear constraint solutions. The resulting contributions include couplings to the other perturbations and a quartic longitudinal interaction,
	\begin{equation}
		\begin{aligned}
			\mathcal L_{\mathrm{int}}\supset{}&
			-\frac{\Mpl^2w_2}{q_S}\,
			\dot\gamma_{ij}\partial_i\pi_c\partial_j\pi_c
			-\frac{2\Mpl^2w_2}{\bar{A}q_S}\,
			\frac{V_i}{a}\partial_i\pi_c\Box\pi_c\\
			&\qquad\qquad-\frac{\sqrt2w_2}{\Mpl\sqrt{q_S}}\,
			\pi_c(\delta\dot\varphi_N)^2
			-\frac{2\Mpl^4w_2^2}{3\bar{A}^2q_S^2}\,
			[(\partial\pi_c)^2]^2.
		\end{aligned}
		\label{eq:mixed-vertices}
	\end{equation}
	After canonically normalising the tensor, the first two coefficients are of order $(g\Mpl^2)^{-1}$, the matter coefficient is of order $\Mpl^{-1}$, and the quartic coefficient is of order $(g^2\Mpl^4)^{-1}$. The first two
	interactions and the quartic term therefore become important at energies of order $\sqrt{g}\Mpl$, while the matter interaction is suppressed by $\Mpl$. For $g\ll1$, these scales lie above the longitudinal cubic scale $g^{2/3}\Mpl$. We therefore have that the longitudinal cubic interaction sets the lowest
	scale among the terms considered here. Combining this estimate with the condition $E\gg g\Mpl$ used in solving the constraints gives the energy range
	\begin{equation}
		g\Mpl\ll E\ll g^{2/3}\Mpl
	\end{equation}
	in which the high-frequency approximation applies and the interactions considered here remain perturbative. These interaction scales are nonzero at both vacuum endpoints. Establishing the corresponding range for the full coupled system requires including the remaining interactions and the contributions from the nonlinear constraints, which we leave for future work.
	
	\section{Conclusions}\label{sec:conclusions}
	We have constructed a geodesically complete bounce in cubic Generalised Proca theory with a free canonical matter scalar obeying the null energy condition. The solution connects contracting and expanding de Sitter regions through a smooth minimum of the scale factor. The temporal vector evolves along its algebraic constraint, and its time dependence permits nonzero derivative mixing at the bounce. This example contributes to the study of vector-supported nonsingular cosmologies and the role of the temporal constraint in their stability \cite{Jimenez2016,Bohnenblust2025}. The scalar gradient condition also contains the intrinsic vector contribution identified in Ref.~\cite{Heisenberg2018}. Although the background field strength vanishes, the electric field of the scalar perturbations contributes to this condition and allows positive gradient coefficients throughout the solution.
	
	For sufficiently small positive $\epsilon$, the kinetic terms are positive at every finite time for nonzero comoving wavenumber, and short-wavelength propagation is stable throughout the history. To follow each mode as its physical wavelength grows near the de Sitter endpoints, we use the full linear equations. Their endpoint solutions, together with regular evolution and explicit reconstruction of the constrained fields through the bounce, establish that local perturbations remain bounded. Our use of perturbation variables that remain regular at the bounce follows the considerations developed for nonsingular cosmologies in Refs.~\cite{Ijjas2018,Heisenberg2018}.
	
	We have also estimated the scales at which selected interactions become important. Studies of nonsingular Horndeski cosmologies have likewise examined the hierarchy between interaction scales and the scales of background evolution \cite{Ageeva2021,Choi2025}. In the present model, the direct longitudinal cubic interaction gives $\Lambda_3\sim g^{2/3}\Mpl$, while the selected mixed and quartic interactions are suppressed by higher scales for $g\ll1$. The resulting local range $g\Mpl\ll E\ll g^{2/3}\Mpl$ lies within the high-frequency approximation and keeps the interactions considered here perturbative. The estimated scales remain nonzero at both vacuum endpoints and exceed the background curvature scales for sufficiently small $g$.
	
	Determining the corresponding range for the full coupled system requires the full set of interactions and the contributions from the nonlinear constraints, which we leave, together with nonlinear stability and a transition from the bounce to a standard radiation era, for future work. This model provides an explicit example in which the background and local linear perturbations can be followed through a geodesically complete cosmological bounce while the matter sector remains canonical and satisfies the null energy condition.
	
	\subsubsection*{Acknowledgements}
	The author was supported by the Institute for Basic Science under project code IBS-R018-D3 and acknowledges support from PSL/Observatoire de Paris. The author also thanks Mohammad Ali Gorji for useful discussions. OpenAI ChatGPT 6 Astra was used to assist with literature search and verification of some analytical relations. All results were independently verified by the author, who takes full responsibility for the scientific content of the manuscript.
	
	\begingroup
	\small
	\bibliographystyle{JHEPmod}
	\bibliography{proca_bounce}

@article{Heisenberg2014,
 author={Heisenberg, Lavinia}, title={Generalization of the {Proca} Action},
 journal={JCAP}, volume={05}, number={2014}, pages={015}, year={2014},
 doi={10.1088/1475-7516/2014/05/015}, eprint={1402.7026}, archivePrefix={arXiv}, primaryClass={hep-th}}

@article{Tasinato2014,
 author={Tasinato, Gianmassimo}, title={Cosmic Acceleration from {Abelian} Symmetry Breaking},
 journal={JHEP}, volume={04}, number={2014}, pages={067}, year={2014},
 doi={10.1007/JHEP04(2014)067}, eprint={1402.6450}, archivePrefix={arXiv}, primaryClass={hep-th}}

@article{Kobayashi2016,
 author={Kobayashi, Tsutomu}, title={Generic instabilities of nonsingular cosmologies in {Horndeski} theory: A no-go theorem},
 journal={Phys. Rev. D}, volume={94}, pages={043511}, year={2016}, doi={10.1103/PhysRevD.94.043511}, eprint={1606.05831}, archivePrefix={arXiv}, primaryClass={hep-th}}

@article{Creminelli2016,
 author={Creminelli, Paolo and Pirtskhalava, David and Santoni, Luca and Trincherini, Enrico},
 title={Stability of Geodesically Complete Cosmologies}, journal={JCAP}, volume={11}, number={2016}, pages={047}, year={2016},
 doi={10.1088/1475-7516/2016/11/047}, eprint={1610.04207}, archivePrefix={arXiv}, primaryClass={hep-th}}

@article{Libanov2016,
 author={Libanov, Mikhail and Mironov, Sergey and Rubakov, Valery},
 title={Generalized {Galileons}: Instabilities of bouncing and {Genesis} cosmologies and modified {Genesis}},
 journal={JCAP}, volume={08}, number={2016}, pages={037}, year={2016},
 doi={10.1088/1475-7516/2016/08/037}, eprint={1605.05992}, archivePrefix={arXiv}, primaryClass={hep-th}}

@article{Jimenez2016,
 author={Beltr{\'a}n Jim{\'e}nez, Jose and Heisenberg, Lavinia and Koivisto, Tomi S.},
 title={Cosmology for quadratic gravity in generalized {Weyl} geometry}, journal={JCAP}, volume={04}, number={2016}, pages={046}, year={2016},
 doi={10.1088/1475-7516/2016/04/046}, eprint={1602.07287}, archivePrefix={arXiv}, primaryClass={hep-th}}

@article{Heisenberg2018,
 author={Heisenberg, Lavinia and Kase, Ryotaro and Tsujikawa, Shinji},
 title={Gauge-ready formulation of cosmological perturbations in scalar-vector-tensor theories},
 journal={Phys. Rev. D}, volume={98}, pages={123504}, year={2018},
 doi={10.1103/PhysRevD.98.123504}, eprint={1807.07202}, archivePrefix={arXiv}, primaryClass={gr-qc}}

@article{Bohnenblust2025,
 author={Bohnenblust, Lara and Giardino, Serena and Heisenberg, Lavinia and Nussbaumer, Nadine},
 title={To bounce or not to bounce in generalized {Proca} theory and beyond}, journal={JHEP}, volume={07}, number={2025}, pages={124}, year={2025},
 doi={10.1007/JHEP07(2025)124}, eprint={2412.03977}, archivePrefix={arXiv}, primaryClass={gr-qc}}

@article{Ijjas2018,
 author={Ijjas, Anna}, title={Space-time slicing in {Horndeski} theories and its implications for non-singular bouncing solutions},
 journal={JCAP}, volume={02}, number={2018}, pages={007}, year={2018},
 doi={10.1088/1475-7516/2018/02/007}, eprint={1710.05990}, archivePrefix={arXiv}, primaryClass={gr-qc}}

@article{Ijjas2016,
 author={Ijjas, Anna and Steinhardt, Paul J.}, title={Classically Stable Nonsingular Cosmological Bounces},
 journal={Phys. Rev. Lett.}, volume={117}, pages={121304}, year={2016},
 doi={10.1103/PhysRevLett.117.121304}, eprint={1606.08880}, archivePrefix={arXiv}, primaryClass={gr-qc}}

@article{Ageeva2020,
 author={Ageeva, Yulia and Petrov, Pavel and Rubakov, Valery}, title={{Horndeski} genesis: Consistency of classical theory},
 journal={JHEP}, volume={12}, number={2020}, pages={107}, year={2020},
 doi={10.1007/JHEP12(2020)107}, eprint={2009.05071}, archivePrefix={arXiv}, primaryClass={hep-th}}

@article{MolinaParis1999,
  author = {Molina-Par{\'i}s, Carmen and Visser, Matt},
  title = {Minimal conditions for the creation of a {Friedman--Robertson--Walker} universe from a ``bounce''},
  journal = {Phys. Lett. B}, volume = {455}, pages = {90--95}, year = {1999},
  doi = {10.1016/S0370-2693(99)00469-4}, eprint = {gr-qc/9810023}, archivePrefix = {arXiv}
}

@article{Ageeva2021,
  author = {Ageeva, Yulia and Petrov, Pavel and Rubakov, Valery},
  title = {Nonsingular cosmological models with strong gravity in the past},
  journal = {Phys. Rev. D}, volume = {104}, pages = {063530}, year = {2021},
  doi = {10.1103/PhysRevD.104.063530}, eprint = {2104.13412}, archivePrefix = {arXiv}, primaryClass = {hep-th}
}

@article{Choi2025,
  author = {Choi, Han Gil and Petrov, Pavel and Yamaguchi, Masahide},
  title = {Can {Horndeski Genesis} be nonpathological?},
  journal = {JHEP}, volume = {08}, pages = {044}, year = {2025},
  doi = {10.1007/JHEP08(2025)044}, eprint = {2503.02626}, archivePrefix = {arXiv}, primaryClass = {hep-th}
}

@article{Novello2008,
  author = {Novello, M. and Perez Bergliaffa, S. E.},
  title = {Bouncing Cosmologies},
  journal = {Phys. Rept.},
  volume = {463},
  pages = {127--213},
  year = {2008},
  doi = {10.1016/j.physrep.2008.04.006},
  eprint = {0802.1634},
  archivePrefix = {arXiv},
  primaryClass = {astro-ph}
}

@article{Battefeld2015,
  author = {Battefeld, Diana and Peter, Patrick},
  title = {A critical review of classical bouncing cosmologies},
  journal = {Phys. Rept.},
  volume = {571},
  pages = {1--66},
  year = {2015},
  doi = {10.1016/j.physrep.2014.12.004},
  eprint = {1406.2790},
  archivePrefix = {arXiv},
  primaryClass = {astro-ph.CO}
}

@article{Khoury2001,
  author = {Khoury, Justin and Ovrut, Burt A. and
            Steinhardt, Paul J. and Turok, Neil},
  title = {The Ekpyrotic Universe: Colliding Branes and the
           Origin of the Hot Big Bang},
  journal = {Phys. Rev. D},
  volume = {64},
  pages = {123522},
  year = {2001},
  doi = {10.1103/PhysRevD.64.123522},
  eprint = {hep-th/0103239},
  archivePrefix = {arXiv}
}

@article{Erickson2004,
  author = {Erickson, Joel K. and Wesley, Daniel H. and
            Steinhardt, Paul J. and Turok, Neil},
  title = {{Kasner} and {Mixmaster} behavior in universes
           with equation of state {$w \geq 1$}},
  journal = {Phys. Rev. D},
  volume = {69},
  pages = {063514},
  year = {2004},
  doi = {10.1103/PhysRevD.69.063514},
  eprint = {hep-th/0312009},
  archivePrefix = {arXiv}
}

@article{Creminelli2010,
  author = {Creminelli, Paolo and Nicolis, Alberto and
            Trincherini, Enrico},
  title = {{Galilean Genesis}: an alternative to inflation},
  journal = {JCAP},
  volume = {11},
  pages = {021},
  year = {2010},
  doi = {10.1088/1475-7516/2010/11/021},
  eprint = {1007.0027},
  archivePrefix = {arXiv},
  primaryClass = {hep-th}
}

@article{Deffayet2010,
  author = {Deffayet, Cedric and Pujolas, Oriol and
            Sawicki, Ignacy and Vikman, Alexander},
  title = {Imperfect Dark Energy from Kinetic Gravity Braiding},
  journal = {JCAP},
  volume = {10},
  pages = {026},
  year = {2010},
  doi = {10.1088/1475-7516/2010/10/026},
  eprint = {1008.0048},
  archivePrefix = {arXiv},
  primaryClass = {hep-th}
}

@article{Qiu2011,
  author = {Qiu, Taotao and Evslin, Jarah and Cai, Yi-Fu and
            Li, Mingzhe and Zhang, Xinmin},
  title = {Bouncing {Galileon} Cosmologies},
  journal = {JCAP},
  volume = {10},
  pages = {036},
  year = {2011},
  doi = {10.1088/1475-7516/2011/10/036},
  eprint = {1108.0593},
  archivePrefix = {arXiv},
  primaryClass = {hep-th}
}

@article{Easson2011,
  author = {Easson, Damien A. and Sawicki, Ignacy and
            Vikman, Alexander},
  title = {{G-Bounce}},
  journal = {JCAP},
  volume = {11},
  pages = {021},
  year = {2011},
  doi = {10.1088/1475-7516/2011/11/021},
  eprint = {1109.1047},
  archivePrefix = {arXiv},
  primaryClass = {hep-th}
}

@article{Koehn2016,
  author = {Koehn, Michael and Lehners, Jean-Luc and Ovrut, Burt},
  title = {Nonsingular bouncing cosmology:
           Consistency of the effective description},
  journal = {Phys. Rev. D},
  volume = {93},
  pages = {103501},
  year = {2016},
  doi = {10.1103/PhysRevD.93.103501},
  eprint = {1512.03807},
  archivePrefix = {arXiv},
  primaryClass = {hep-th}
}

@article{Cai2017,
  author = {Cai, Yong and Wan, Youping and Li, Hai-Guang and
            Qiu, Taotao and Piao, Yun-Song},
  title = {The Effective Field Theory of nonsingular cosmology},
  journal = {JHEP},
  volume = {01},
  pages = {090},
  year = {2017},
  doi = {10.1007/JHEP01(2017)090},
  eprint = {1610.03400},
  archivePrefix = {arXiv},
  primaryClass = {gr-qc}
}

@article{Kolevatov2017,
  author = {Kolevatov, R. and Mironov, S. and Sukhov, N. and
            Volkova, V.},
  title = {Cosmological bounce and {Genesis} beyond {Horndeski}},
  journal = {JCAP},
  volume = {08},
  pages = {038},
  year = {2017},
  doi = {10.1088/1475-7516/2017/08/038},
  eprint = {1705.06626},
  archivePrefix = {arXiv},
  primaryClass = {hep-th}
}

@article{Mironov2018,
  author = {Mironov, S. and Rubakov, V. and Volkova, V.},
  title = {Bounce beyond {Horndeski} with {GR} asymptotics
           and {$\gamma$}-crossing},
  journal = {JCAP},
  volume = {10},
  pages = {050},
  year = {2018},
  doi = {10.1088/1475-7516/2018/10/050},
  eprint = {1807.08361},
  archivePrefix = {arXiv},
  primaryClass = {hep-th}
}

@article{Allys2016,
  author = {Allys, Erwan and Peter, Patrick and Rodriguez, Yeinzon},
  title = {Generalized {Proca} action for an {Abelian} vector field},
  journal = {JCAP},
  volume = {02},
  pages = {004},
  year = {2016},
  doi = {10.1088/1475-7516/2016/02/004},
  eprint = {1511.03101},
  archivePrefix = {arXiv},
  primaryClass = {hep-th}
}

@article{DeFelicePerturbations2016,
  author = {De Felice, Antonio and Heisenberg, Lavinia and
            Kase, Ryotaro and Mukohyama, Shinji and
            Tsujikawa, Shinji and Zhang, Ying-li},
  title = {Effective gravitational couplings for cosmological
           perturbations in generalized {Proca} theories},
  journal = {Phys. Rev. D},
  volume = {94},
  pages = {044024},
  year = {2016},
  doi = {10.1103/PhysRevD.94.044024},
  eprint = {1605.05066},
  archivePrefix = {arXiv},
  primaryClass = {gr-qc}
}

@article{Horndeski1974,
  author = {Horndeski, Gregory Walter},
  title = {Second-order scalar-tensor field equations
           in a four-dimensional space},
  journal = {Int. J. Theor. Phys.},
  volume = {10},
  number = {6},
  pages = {363--384},
  year = {1974},
  doi = {10.1007/BF01807638}
}
	\endgroup
	
\end{document}